\documentstyle{article}\begin{document}
\title{Stabilization of Starobinsky-Vilenkin stochastic inflation by an environmental noise }
\author{ Z. Haba\\
Institute of Theoretical Physics, University of Wroclaw,\\ 50-204
Wroclaw, Plac Maxa Borna 9, Poland\\
email:zhab@ift.uni.wroc.pl}\maketitle
\begin{abstract} We discuss the inflaton $\phi$ in an interaction
with an infinite number of fields treated as a random environment
(noise) with a friction $\gamma^{2}>0$. In a Markovian
approximation we obtain a stochastic wave equation (appearing also
in the warm inflation models). After the replacement of the
environment by the white noise the  stochastic wave equation
violates the energy conservation if $\gamma\neq 0$. We introduce a
dark energy as a compensation of the inflaton energy-momentum.
 We add to the classical wave equation the Starobinsky-Vilenkin
 noise which in the slow-roll approximation
describes the quantum fluctuations in an expanding metric. We
investigate the resulting consistent stochastic
Einstein-Klein-Gordon system in the slow-roll regime. We obtain
Fokker-Planck equation for the probability distribution of the
inflaton assuming that the dependence of the scale factor $a$ and
the Hubble variable $H$ on the field $\phi$ is known. We obtain
explicit stationary solutions of the Fokker-Planck equation
assuming that $a(\phi)$ and $H(\phi)$ can approximately be
determined in a slow roll regime with a neglect of noise. We
extend the results to the multifield $D$-dimensional configuration
space. We show that in the regime
$a(\phi)^{3}H(\phi)^{5}\rightarrow \infty$ the quantum noise
determines the asymptotic behaviour of the stationary
distribution. If $a^{3}H^{5}$ stays finite then the environmental
noise ensures the integrability of the stationary probability. In
such a case there is no need to introduce boundary conditions with
the purpose to eliminate infinite inflation. The variation of
$a^{3}H^{5}$ could be interpreted as a sign of a transition from
cold inflation to warm inflation.

\end{abstract}

\section{Introduction}The inflationary model agrees very well with the observational data \cite{ade}.
However, the model is a compilation of classical and quantum field
theory. We consider an interaction of a classical scalar field
with gravity in order to generate inflation and to derive the
$\Lambda$CDM model. Then, quantization of scalar  and tensor
fields (linearized gravity) is introduced in order to explain the
observed CMB fluctuations. These fluctuations agree  with the
observed CMB power spectrum (there are some difficulties with low
multipoles). Starobinsky \cite{star} \cite{starobinsky} and
Vilenkin \cite{vilenkin} suggested an approach which treats
quantum scalar field non-perturbatively in classical Einstein
equations. Their idea is based on the observation that the quantum
scalar field in an expanding universe behaves as a classical
diffusion process. In such a case we obtain a stochastic
Einstein-Klein-Gordon (EKG)system. In the slow-roll limit the
stochastic EKG system is approximated by a diffusion equation.

In some approaches to cosmology it is pointed out that it is a
paramount simplification to assume that there is only one field
$\phi$ (inflaton) interacting with gravity. The inflaton as well
as the observed luminous matter may be interacting with an
infinite set of quantum fields creating particles which are not
observed (this is consistent with the string theory point of view
\cite{string}\cite{string2}). Each field gives a negligible
contribution to the interaction but the infinite number of these
fields forms an environment which can be treated by methods of
statistical mechanics (central limit theorem) in a close analogy
to the motion of a Brownian particle in a fluid
\cite{fordkac}\cite{ford}. As a result we obtain a stochastic wave
equation for an inflaton with two perturbations: noise from the
environment (thermal noise) and the quantum noise. The stochastic
wave equation  determines correlation functions which are measured
in astronomical observations ,e.g.,
 the power
spectrum of fluctuations. In such a  model one can ask questions
intrinsic to a quantum theory, e.g., the probability of an
appearance of the universe of a given size or  the probability
distribution of the inflaton \cite{linde}. The
Starobinsky-Vilenkin theory gives an analytic formula for this
probability at large time (stationary solution). The same result
follows from quantization of gravity either on the basis of
deWitt-Wheeler equation \cite{dewitt} or using the path-integral
\cite{hh}. However, for some potentials the resulting stationary
distribution is not normalizable. The result applies well in the
range of slow-roll which coincides with the range of the field
$\phi$ in the course of inflation. However, in order to restrict
the range of evolution of $\phi$ in the stochastic equation we  must
introduce  boundary
conditions as is done in \cite{venn}\cite{venn3}. The theory of
diffusions with boundary conditions is more involved \cite{ikeda}
and no explicit estimates on the stationary probability
distribution are available. We show in this paper that for some
potentials which require boundary conditions in the approach
\cite{venn}-\cite{venn3} the environmental noise cures the
integrability problem. In such a case the environmental noise could be treated as an alternative
 regularization procedure which gives some sense to the stationary evolution beyond the inflationary regime.
 We suggest  the following interpretation:
when the cold stochastic inflation of Starobinsky is terminating
the warm inflation excited by the environmental noise dominates
leading to a smooth behaviour at large time

In this paper we treat the stochastic wave equation with friction $\gamma>0$  and with thermal and
quantum noise as an approximation to the Einstein-Klein-Gordon
(EKG) equations. In such a case the energy-momentum tensor must be
conserved. The noise in the inflaton equation violates the
conservation law if $\gamma\neq 0$ . We add to the energy-momentum of the inflaton
$\phi$ a correction which ensures the conservation law and can be
associated with the energy-momentum of the dark energy ( the idea
expressed also in \cite{sudarsky}). In such a case we can insert
the stochastic field in Einstein equations arriving at stochastic
Einstein equations.  In a homogeneous flat universe we obtain a
closed system of stochastic Friedmann equations. This system of
equations determines the Fokker-Planck equation for the
probability distribution. It contains many variables. In order to
solve this equation we make some simplifying assumptions. The main
approximation involves the assumption that the Hubble parameter
$H(\phi)$ and the scale factor $a(\phi)$ can be expressed from the
slow-roll no noise approximation of the EKG system. We show that
the Fokker-Planck equation in a limit of small
$a(\phi)^{3}H(\phi)^{5}$ has a stationary solution which has the
same asymptotics as the solution we would have obtained solely
from the thermal noise .  For large $a^{3}H^{5}$  the asymptotics
of the solution derived by Starobinsky  (the Hartle-Hawking
probability density) results. The transition from large
$a^{3}H^{5}$ to small $a^{3}H^{5}$ could be interpreted as a
transition from cold inflation to warm inflation. The probability
distribution of the inflaton gives a non-perturbative formula for
quantum fluctuations with the thermal noise. Berera et
al\cite{bererafang} \cite{warm} \cite{power1}\cite{power2}
calculate the power spectrum using the stochastic equation with
the thermal noise (warm inflation). In \cite{starven}      and
\cite{venn} the power spectrum is calculated using the
Starobinsky-Vilenkin approach to the cold inflation. There are
some arguments \cite{stochmulti} that the non-perturbative
treatment of fluctuations can explain the low multipoles discord.
 From the calculations in \cite{starven}\cite{venn}  \cite{power1}\cite{power2} \cite{stochmulti} it can be
seen that the addition  of the thermal noise to cold inflation
will change the power spectrum (we calculate it in a subsequent
paper \cite{habaejc}).

We discuss a general framework of Einstein gravity with stochastic
wave equations in secs.2-3. In sec.4 we make the slow-roll
approximation for stochastic equations. We derive the
Fokker-Planck equation assuming that the functions $a(\phi)$ and
$H(\phi)$ are known. In sec.5 we obtain  approximate formulas
for these functions  neglecting their noise dependence. We
discuss the stationary solutions of the Fokker-Planck equation in
sec.6.
\section{The energy-momentum tensor of the
interaction with a random environment}
 We consider
an infinite set of scalar fields $\chi^{b}$ of mass $m_{b}$
 coupling to the inflaton with an interaction Lagrangian ${\cal L}_{I}=\sum_{b}\lambda_{b}\chi^{b}\phi$
 ( an  outline of this scheme  has been presented in \cite{habaacta}, detailed calculations  appeared in \cite{habaprep}).
We solve the evolution equations for $\chi^{b}$ with the initial
conditions whose probability distribution is defined by classical
or quantum Gibbs distribution. In a Markovian approximation,
assuming that
$\gamma^{2}\simeq\sum_{b}\lambda_{b}^{2}m_{b}^{-2}<\infty$ and
eliminating the fields $\chi^{b}$ from the equation of motion of
the inflaton, we obtain in a flat expanding metric
\begin{displaymath}
ds^{2}=g_{\mu\nu}dx^{\mu}dx^{\nu}=dt^{2}-a^{2}d{\bf x}^{2}
\end{displaymath} a stochastic wave equation of the form (this equation
has been  derived earlier in \cite{berrera})
\begin{equation}
\partial_{t}^{2}\phi-a^{-2}\triangle\phi+(3H+\gamma^{2})\partial_{t}\phi+m^{2}\phi+V^{\prime}(\phi)+\frac{3}{2}\gamma^{2}H\phi=\eta
 \end{equation}
where $H=a^{-1}\partial_{t}a$ and  $\eta$ is a noise.

In an expanding universe the quantum field behaves like a
classical diffusion process \cite{starobinsky}\cite{vilenkin}. We
can take into account quantum effects by an addition of the
Starobinsky-Vilenkin noise to the rhs of eq.(1). Now,
\begin{equation}
\eta\equiv \partial_{t}\xi=\gamma
a^{-\frac{3}{2}}\partial_{t}B+\frac{3}{2\pi}H^{\frac{5}{2}}\partial_{t}W
\end{equation}
where the first term describes the environmental noise and the
second term the quantum noise. The factor $a^{-\frac{3}{2}}$ comes
from $\det\vert g_{\mu\nu}\vert ^{\frac{1}{4}}$ and the factor
$H^{\frac{5}{2}}$ is chosen in order to reproduce the correlation
functions of the quantum scalar field in an expanding universe
\cite{starobquan}. The (white) noise $\partial_{t}{B}$ is the the
Gaussian random process with the covariance
\begin{equation}
\langle
\partial_{t}B\partial_{s}B\rangle=\delta(t-s)\end{equation}
$\partial_{t}W$ is an independent Gaussian stochastic process with
the same covariance (3).

The  energy-momentum tensor of the scalar field in the presence of
friction $\gamma^{2}\partial_{t}\phi$ and noise $\xi$ is not
conserved (as far as we know this point has not been discussed in
papers on stochastic inflation) . We have to compensate the
energy-momentum by means of a compensating energy-momentum
$T_{de}$ which we associate with the dark sector .
 Now, the conserved energy-momentum tensor
$T_{tot}^{\mu\nu}$ is
\begin{displaymath} T_{tot}^{\mu\nu}=T^{\mu\nu}+T^{\mu\nu}_{de},\end{displaymath}
where $T_{de}$ denotes some other constituents of the primeval
universe. From the conservation law
\begin{equation}
(T_{de}^{\mu\nu})_{;\mu}=-(T^{\mu\nu})_{;\mu}.\end{equation}
 We assume  the energy-momentum in the form of
 an ideal fluid
\begin{equation}
T^{\mu\nu}_{de}=(\rho_{de}+p_{de})u^{\mu}u^{\nu}-g^{\mu\nu}p_{de},
\end{equation}where $\rho$ is the energy density and $p$ is the
pressure.  The velocity $u^{\mu}$ satisfies the normalization
condition
\begin{displaymath}  g_{\mu\nu}u^{\mu}u^{\nu}=1.\end{displaymath}

For the scalar field we have the representation (5) with
\begin{equation}
u^{\mu}=\partial^{\mu}\phi(\partial^{\sigma}\phi\partial_{\sigma}\phi)^{-\frac{1}{2}},
\end{equation}
\begin{equation}
\rho+p=\partial^{\sigma}\phi\partial_{\sigma}\phi,
\end{equation}
\begin{equation}p=\frac{1}{2}\partial^{\sigma}\phi\partial_{\sigma}\phi-V.
\end{equation}
 If $\phi$ satisfies eq.(1) then we have for the scalar field (6)-(8)
\begin{equation}
(T^{\mu\nu})_{;\mu}=\partial^{\nu}\phi(\eta-\gamma^{2}\partial_{t}\phi-\frac{3}{2}\gamma^{2}H\phi)
.
\end{equation}
We interpret eq.(9) as a stochastic equation in the Stratonowitch
sense \cite{ikeda}. The Stratonowitch differential is the only one
which satisfies the Leibniz rule.The divergence equation (9) for
$T^{0\nu}$ in a homogeneous metric in the frame $u=(1,{\bf 0})$
(spatial homogeneity of $\phi$) gives (we apply the stochastic
calculus and the $\circ$-circle notation for stochastic
multiplication of ref. \cite{ikeda}  )
\begin{equation}\begin{array}{l} d\rho+3(1+w_{I})H\rho dt=
\partial_{t}\phi\circ d\xi -\frac{3}{2}\gamma^{2}H\phi\partial_{t}\phi dt-\gamma^{2}(\partial_{t}\phi)^{2}dt
,\end{array}\end{equation}
  For a
scalar field with potential $V$ we have
\begin{equation} w_{I}=
(\frac{1}{2}(\partial_{t}\phi)^{2}-V)(\frac{1}{2}(\partial_{t}\phi)^{2}+V)^{-1}.
\end{equation} According to eq.(4) the compensating energy density must have the (non)conservation
law with an opposite sign
\begin{equation}\begin{array}{l}
d\rho_{de}+3H(1+w)\rho_{de}dt=\frac{3}{2}\gamma^{2}H\phi\partial_{t}\phi
dt+\gamma^{2}(\partial_{t}\phi)^{2}dt-\partial_{t}\phi \circ
d\xi,\end{array}
\end{equation}where
\begin{equation}
w=\frac{p_{de}}{\rho_{de}}\end{equation} is not determined by
eq.(4). If a concrete model of the environmental scalar fields
$\chi_{b}$ (discussed at the beginning of this section) is chosen
then $w$ would be determined by this model. In order to have  a
closed system of equations an approximation $w\simeq const$ would
be useful. For scalar fields $p+\rho=(1+w)\rho\simeq 0$ if the
kinetic energy of the $\chi_{b}$ fields is negligible. We choose
the interaction potential for the $\chi_{b}$ fields
\cite{habaprep} which justifies the approximation $w\simeq -1$.

We have obtained in eqs.(10)-(12) a special form of the energy
non-conservation. Some other  models with a non-zero term on the
rhs of eq.(10) (interpreted as a time derivative of the
cosmological term) have been discussed
\cite{review}\cite{sola}\cite{bas}\cite{ss}. In  models of
inflation \cite{steinhardt}\cite{barerarad}\cite{bererafang}  the
term $\gamma^{2}(\partial_{t}\phi)^{2}$ describes a transformation
of the inflaton energy into radiation. In such a case we add
$\rho_{rad}$ to the total energy-momentum, omit the term
$\gamma^{2}(\partial_{t}\phi)^{2}$ in the equation (12) for dark
energy and in order to preserve the energy conservation we assume
\begin{displaymath}
\partial_{t}\rho_{rad}+4\rho_{rad}=\gamma^{2}(\partial_{t}\phi)^{2}.
\end{displaymath}
We insert the conserved energy-momentum in Einstein equations
\begin{equation}
 R^{\mu\nu}-\frac{1}{2}g^{\mu\nu}R=8\pi G
T_{tot}^{\mu\nu},
\end{equation} where $R^{\mu\nu}$ is the Ricci tensor and $G$ is the Newton constant.
The Friedman equation in the FRLW flat metric  reads
\begin{equation}\begin{array}{l}
H^{2}=\frac{8\pi G}{3}(\rho+\rho_{de}).\end{array}\end{equation}
We differentiate eq.(15). Then, together with the stochastic
equations (10) and (12) for the densities $\rho$, we have a closed
system of stochastic equations
\begin{equation}
da=Hadt
\end{equation}
\begin{equation}
dH^{2}=\frac{8\pi G}{3}(d\rho+d\rho_{de}) =-8\pi
GH\Big((1+w_{I})\rho+(1+w)\rho_{de}\Big)dt\end{equation} We can
insert in eq.(17) the solution of eqs.(10),(12) for $\rho$ as a
function of $a$ and subsequently insert $H$ as a function of $a$
obtaining a stochastic integro-differential equation (16) for $a$.
We choose our model of $\chi_{b}$ fields in such a way that
$w\simeq-1$ in eqs.(12) and (17). Then, eq.(10) determines $\rho$
whereas eq.(17) does not depend explicitly on the dark energy.
Eqs.(1),(7)-(8) and (16)-(17) (with $w=-1$) constitute a closed
system of differential equations involving $\phi,H,a$ and the
noise $\eta$.

\section{Stochastic Einstein-Klein-Gordon system}
 Eq.(17) does not depend on the details of the
compensating matter if $w=-1$. In such a case we obtain a closed
random dynamical system (in the first order form) with the
Starobinsky-Vilenkin noise $W$ and the environmental noise  $B$
\begin{equation} d\phi=\Pi dt ,
\end{equation}
\begin{equation}
d\Pi=-(3H+\gamma^{2})\Pi dt -V^{\prime}dt
-\frac{3}{2}\gamma^{2}H\phi dt+\gamma a^{-\frac{3}{2}}\circ
dB+\frac{3}{2\pi}H^{\frac{5}{2}}\circ dW  ,
\end{equation}
\begin{equation}
dH=-4\pi G\Pi^{2}dt   ,
\end{equation}\begin{equation}
da=H adt .
\end{equation}
For a comparison with  ref.\cite{starven} and a subsequent study of
stochastic inflation it will be useful to change the world time
$t$ into the e-folding time $\nu$ describing the change of the
scale factor
\begin{equation}
\nu=\int_{0}^{t}Hds=\ln(\frac{a}{a_{0}}) .
\end{equation}
In general relativity there is a freedom in the choice of the time
variable. The e-folding time allows to integrate some stochastic
equations \cite{stochmulti}\cite{habaejc}. This is  also a proper
choice of time for a description of cosmological perturbations and
to calculate the power spectrum\cite{starven}. Then (for a simple
description of a stochastic change of time see \cite{gikhman})
\begin{equation}
d\phi=H^{-1}\Pi d\nu ,
\end{equation}
\begin{equation}\begin{array}{l}
d\Pi=-(3H+\gamma^{2})H^{-1}\Pi d\nu -V^{\prime}H^{-1}d\nu
-\frac{3}{2}\gamma^{2}\phi d\nu\cr+\gamma
\exp(-\frac{3}{2}\nu)H^{-\frac{1}{2}}\circ
dB(\nu)+\frac{3}{2\pi}H^{2}\circ dW(\nu),\end{array}
\end{equation}
\begin{equation}
dH=-4\pi G\Pi^{2}H^{-1}d\nu.
\end{equation}
The stochastic wave equation  leads to a definite relation between
$\phi,\Pi, H $ and $a$. We multiply eq.(19) by $\Pi$ and using
(20) we obtain the following identity (which could be considered
as an equation for $H(\phi,\Pi,B,W)$)
\begin{equation}\begin{array}{l}
\partial_{t}(\frac{1}{2}\Pi^{2}+V-\frac{3}{4}\gamma^{2}H\phi^{2}-\frac{3}{8\pi
G}H^{2}+\frac{1}{4\pi G}\gamma^{2}H)-3\pi
G\gamma^{2}\phi^{2}\Pi^{2}\cr= \gamma
a^{-\frac{3}{2}}\Pi\circ\partial_{t}B+\frac{3}{2\pi}H^{\frac{5}{2}}\Pi\circ\partial_{t}W
. \end{array}\end{equation} From eqs.(24)-(25) we obtain a similar
equation for the $\nu$-derivative
\begin{equation}\begin{array}{l}
\partial_{\nu}(\frac{1}{2}\Pi^{2}+V-\frac{3}{4}\gamma^{2}H\phi^{2}-\frac{3}{8\pi
G}H^{2}+\frac{1}{4\pi G}\gamma^{2}H)-3\pi
G\gamma^{2}H^{-1}\phi^{2}\Pi^{2}\cr= \gamma
\exp(-\frac{3}{2}\nu)H^{-\frac{1}{2}}\Pi\circ\partial_{\nu}B+\frac{3}{2\pi}H^{2}\Pi\circ\partial_{\nu}W
\end{array}\end{equation}
Integrating eq.(27)
\begin{equation}\begin{array}{l}
\frac{1}{2}\Pi^{2}+V-\frac{3}{4}\gamma^{2}H\phi^{2}-\frac{3}{8\pi
G}H^{2}+\frac{1}{4\pi G}\gamma^{2}H+V_{0}\cr=3\pi
G\gamma^{2}\int_{0}^{\nu}dsH^{-1}\phi^{2}\Pi^{2}+
\int_{0}^{\nu}ds(\gamma \exp(-\frac{3}{2}s)H^{-1}\Pi\circ
dB+\frac{3}{2\pi}H^{2}\Pi\circ dW) ,\end{array}
\end{equation}(where $V_{0}$ is an arbitrary integration constant related to the cosmological constant $\Lambda$ by $8\pi G V_{0}=\Lambda$ ).
Eqs.(26)-(27) can be considered as  equations for $H$. We can get a solution of
 eq.(27) as a perturbation series in $\gamma$. At $\gamma=0$
eq.(27) has the solution
\begin{equation}
\begin{array}{l}
(\frac{3}{8\pi
G}H^{2}-\frac{1}{2}\Pi^{2}-V-V_{0})(\nu)\cr=\Big(\frac{3}{8\pi
G}H^{2}-\frac{1}{2}\Pi^{2}-V-V_{0}\Big)(0)\exp(-4G\int_{0}^{\nu}\Pi(s)\circ
dW(s))\cr
-4G\int_{0}^{\nu}(\frac{1}{2}\Pi^{2}+V+V_{0})\exp(-4G\int_{s}^{\nu}\Pi(\tau)\circ
dW(\tau))\Pi(s)\circ dW(s),
\end{array}\end{equation}where the first term on the rhs
corresponds to the initial condition for the lhs. Inserting the
expansion $H_{\gamma}=H+\gamma H_{1}+...$ in eq.(28) we obtain
$H(\phi,\Pi,W,B)$ as a power series in $\gamma$. Then,
eqs.(23)-(24) define a non-linear stochastic wave equation
 for $(\phi,\Pi)$. The solution allows to
calculate the probability distribution of $(\phi,\Pi)$.

\section{Stochastic equations for  slow-roll
inflation} The program outlined at the end of  sec.3 to solve the
stochastic Einstein-Klein-Gordon (EKG) equation and calculate the
probability distribution is difficult to perform in a
non-perturbative way. We shall rely on approximations. First, we
neglect the noise in eqs.(27)-(28) and the term
$\gamma^{2}\Pi^{2}\phi^{2}$. Then,
\begin{equation}
H=\gamma^{2}(\frac{1}{3}-4\pi G\phi^{2}) + \sqrt{\frac{8\pi
G}{3}(V+V_{0} )+\frac{1}{2}\Pi^{2}+\gamma^{4}(\frac{1}{3}-4\pi
G\phi^{2})^{2}}.\end{equation} Using, eq.(30) we can obtain the
Fokker-Planck equation for the probability distribution of
$(\phi,\Pi)$ but it would be difficult to find solutions of this
equation. We make the next simplifying assumption $\Pi\simeq 0$
and $\gamma=0$ in eq.(30). Then, the stochastic system (19)-(21) is

\begin{equation}
3Hd\phi = -V^{\prime}dt +\gamma
a^{-\frac{3}{2}}\circ dB+\frac{3}{2\pi}H^{\frac{5}{2}}\circ dW,
\end{equation}
\begin{equation}
dH=-4\pi G(\partial_{t}\phi)^{2}dt ,
\end{equation}\begin{equation}
da=H adt  .
\end{equation}
In the e-fold time the $\Pi\simeq 0$ limit reads (we express $a$
as a function of $\phi$)\begin{equation} 3Hd\phi =
-V^{\prime}H^{-1}d\nu +\gamma a^{-\frac{3}{2}}
H^{-\frac{1}{2}}\circ dB(\nu)+\frac{3}{2\pi}H^{2}\circ dW(\nu) ,
\end{equation}
\begin{equation}
d\ln(H)=-4\pi G(\partial_{\nu}\phi)^{2}d\nu .
\end{equation}The approximation $\Pi\simeq 0$ together with
\begin{equation}
H=\sqrt{\frac{8\pi G}{3}(V+V_{0}
)}.\end{equation}
 is known as the slow-roll approximation. Note that eq.(36) follows from
 eq.(29) and the assumption $\Pi\simeq 0$.  The
slow-roll approximation fails if the variables
$\epsilon=\frac{1}{16\pi G} (V^{\prime})^{2}(V+V_{0})^{-2}$ and
$\tilde{\eta}=\frac{1}{8\pi G}V^{\prime\prime}(V+V_{0})^{-1} $ are
of order 1. For deterministic systems  we can restrict the initial
values of the field and the time evolution in order to satisfy the
requirement of small $\epsilon$ and $\tilde{\eta}$. It is  more
difficult to do it in a random system because the noise can move
the system to the forbidden region of the field configurations. We
can define the stochastic system in a required domain of field
configurations by an imposition of boundary conditions (as
discussed in \cite{venn}\cite{venn3}for the Starobinsky-Vilenkin
stochastic equation with $\gamma=0$). However, the stochastic
process and its probability distribution depend on the boundary
conditions , e.g., for some potentials $V$ the limit $t\rightarrow
\infty$ (the stationary distribution) exists for the process with
the boundary conditions but does not exist if the boundary
conditions are removed. Nevertheless,  some correlation functions
may have a negligible dependence on boundary conditions as
discussed in \cite{venn}\cite{venn3} (for more on boundaries in
diffusions see \cite{ikeda}). We show in the next section that
 in some models of inflation the assumption of $\gamma\neq 0$
allows to work without boundary conditions.

The formula (36) for $H$ has been applied by most  authors on
stochastic inflation. It can be considered as a definition of $V$
in the  Hamiltonian-Jacobi framework
\cite{hamjac}\cite{hamjac2}\cite{pot}. It can be shown that
eq.(36) is exact in a deterministic EKG system
at large time \cite{randal}. If $H$ is known then in principle we
can determine $a(\phi)$ (from eq.(33)) after solving the
stochastic equations. We can obtain an explicit formula if we
neglect the noise (as in eq.(36)) and  apply the slow roll
approximation (31) (without noise). Then,
\begin{equation}
\ln (a)=\int Hdt=\int d\phi (\frac{d\phi}{dt})^{-1}H =-8\pi G
 \int d\phi (V+V_{0})(V^{\prime})^{-1}.
 \end{equation} As discussed in sec.3 the correction
 to eq.(37) could be calculated as a perturbation series in the noise.
Eq.(37) can be derived also from the exact formula for the mean value of the e-folds
as the leading term of the saddle point expansion \cite{starven}\cite{habaejc}.

We can generalize the stochastic equations (31) to multiple scalar
fields \cite{starobmulti}\cite{venn3}$\phi=(\phi^{1},
...,\phi^{D})$. Then, the noise $\eta=(\eta^{1},....,\eta^{D})$
(eq.(2)) consists of independent random Gaussian variables with
the same variance. In eq.(31) $\phi$ and the noises should be
treated as vectors, $V^{\prime}\rightarrow \nabla V$ and
$(\partial_{t}\phi)^{2}\rightarrow \vert
\partial_{t}\phi\vert^{2}$ (the length of the vector $\partial_{t}\phi$). We can calculate $a(\phi)$ if
$V=V(\vert\phi\vert)$ is rotation invariant with
\begin{displaymath}
\vert\phi\vert^{2}=(\phi^{1})^{2}+...+(\phi^{D})^{2}
\end{displaymath}
  Then, (after an omission of noise in eq.(31))
\begin{equation}\ln (a)=\int Hdt=\int d\vert\phi\vert
(\frac{d\vert\phi\vert}{dt})^{-1}H =-8\pi G
 \int d\vert\phi\vert (V+V_{0})\Big(\frac{dV}{d\vert\phi\vert}\Big)^{-1}
 \end{equation}

 For the system
(31)-(33) we have the Fokker-Planck equation \cite{risken} for the
probability distribution of $\phi$ ( we assume the Stratonovich
interpretation of the stochastic equations \cite{ikeda})
\begin{equation}\begin{array}{l}
\partial_{t}P=\frac{\gamma^{2}}{18}\partial_{\phi}\frac{1}{Ha^{\frac{3}{2}}}\partial_{\phi}\frac{1}{Ha^{\frac{3}{2}}}P
+\frac{1}{8\pi^{2}}\partial_{\phi}H^{\frac{3}{2}}\partial_{\phi}H^{\frac{3}{2}}P
+\partial_{\phi}(3H)^{-1}V^{\prime}P .\end{array}\end{equation} In
the e-folding time  eq.(34) determines the probability
distribution as a solution of the equation
\begin{equation}\begin{array}{l}
\partial_{\nu}P=\frac{\gamma^{2}}{18}\partial_{\phi}\frac{1}{H^{\frac{3}{2}}a^{\frac{3}{2}}}\partial_{\phi}\frac{1}{H^{\frac{3}{2}}a^{\frac{3}{2}}}P
+\frac{1}{8\pi^{2}}\partial_{\phi}H\partial_{\phi}H P
+\partial_{\phi}(3H^{2})^{-1}V^{\prime}P\end{array}\end{equation}
In the multifield case ($\phi\in R^{D}$) with
$\partial_{j}=\frac{\partial}{\partial\phi^{j}}$  eq.(39) reads
\begin{equation}\begin{array}{l}
\partial_{t}P=\sum_{j}\partial_{j}\Big(\frac{\gamma^{2}}{18}\frac{1}{Ha^{\frac{3}{2}}}\partial_{j}\frac{1}{Ha^{\frac{3}{2}}}P
+\frac{1}{8\pi^{2}}H^{\frac{3}{2}}\partial_{j}H^{\frac{3}{2}}P
+(3H)^{-1}\partial_{j}V\Big) P .\end{array}\end{equation}

\section{The evolution of the scale factor  $a(\phi)$ in some inflationary
models} We express  $H(\phi)$  as a function of $\phi$ from
eq.(36). The dependence of $a$ on $\phi$ is more involved. We
determine it from eq.(37) obtained  in the slow-roll approximation and no noise limit. All the formulas in
this section can be generalized to a multifield case (38) just by
a replacement $\phi\rightarrow\vert\phi\vert$. Let us consider
some examples of potentials appearing in inflation models
\cite{string2}\cite{pot}\cite{ency}\cite{odintsov}\cite{peeblesratra}\cite{peebles}.
 If $V=\frac{m^{2}}{2}\phi^{2}$
( a chaotic inflation \cite{lindechaotic} ) then from eq.(37)
(the cosmological constant $\Lambda=8\pi G V_{0}$)
\begin{equation}
a=\exp\Big(-4\pi GV_{0} m^{-2}\ln\phi^{2}-2\pi
G\phi^{2}\Big).
\end{equation}
Large $\phi$ corresponds to small $ a$ and small $\phi$ to large
$a$.
 If
$V=g\phi^{n}$ ($n>2$) then
\begin{equation}
a=\exp\Big(-\frac{8\pi GV_{0}}{(2-n)ng}\phi^{2-n}-4\pi
Gn^{-1}\phi^{2}\Big).
\end{equation}
If $\phi\rightarrow \infty$ then $a\rightarrow 0$, if
$\phi\rightarrow 0$ then $a\rightarrow \infty$ (for $\Lambda>0$).

If $V=g\exp(\lambda\phi)$ then
\begin{equation}
a=\exp\Big(\frac{8\pi G
V_{0}}{g\lambda^{2}}\exp(-\lambda\phi)-\frac{8\pi
G}{\lambda}\phi\Big).\end{equation}

If $\phi\rightarrow +\infty$ then $a\rightarrow 0$, if
$\phi\rightarrow -\infty$ then $a\rightarrow \infty$ .

For a flat (plateau) potential \cite{string2}\cite{ency}
\begin{equation}
V=\frac{L+\phi^{2}}{K+\phi^{2}}
\end{equation}
we have
\begin{equation}\begin{array}{l}
a=\exp\Big(-8\pi G\Big(\frac{V_{0}
K^{2}}{4(K-L)}\ln\phi^{2}+\frac{V_{0}
K}{2(K-L)}\phi^{2}+\frac{V_{0} }{8(K-L)}\phi^{4}+\frac{KL
}{4(K-L)}\ln\phi^{2} \cr +
\frac{K+L}{4(K-L)}\phi^{2}+\frac{1}{8(K-L)}\phi^{4}\Big)\Big).\end{array}\end{equation}
If $K>L\geq 0$ then $a\rightarrow 0$ if $\phi\rightarrow \infty$
and $a\rightarrow \infty$ if $\phi\rightarrow 0$. If $L>K>0$ then
$a\rightarrow \infty$ if $\phi\rightarrow \infty$ and
$a\rightarrow 0$ if $\phi\rightarrow 0$ . If $0<K<L$ then
$V^{\prime}< 0$ and
 this is the reason why $a$ is increasing as a function of $\phi$.

Let $V=g\cos\phi$ then
\begin{displaymath}
a=\exp\Big(-8\pi
G\Big(-g^{-1}V_{0}\ln\vert\tan(\frac{\phi}{2})\vert-\ln
\vert\sin(\phi)\vert\Big)\Big).
\end{displaymath} $a\rightarrow 0$
when $\phi\rightarrow 0$.
 When $\phi\rightarrow
\pi$ then $a$ may go to $\infty$ if $g^{-1}\Lambda$ is large
enough (otherwise $a\rightarrow 0$).

The special case
\begin{equation}
V=\vert g\vert (1-\cos\phi)
\end{equation}
corresponds
 to the  " natural inflation" \cite{natural} describing
the axion inflation . From eq.(37) we obtain
\begin{displaymath}
a=\exp\Big(8\pi G\ln\Big(2\cos^{2}(\frac{\phi}{2})\Big)\Big)  .
\end{displaymath}
Here, $-\pi\leq \phi\leq\pi$. From the classical dynamics $\vert \phi\vert$ is decreasing in time and
$\vert\phi\vert\rightarrow 0$ for large time. So, $\phi$ tends to
the minimum of the potential achieving a maximal (finite) value of $a$ at the minimum of the potential.

 For the double well potential
\begin{equation}
V(\phi)=\frac{g}{4}\phi^{4}-\frac{\mu^{2}}{2}\phi^{2}=\frac{g}{4}(\phi^{2}-\frac{\mu^{2}}{g})^{2}-\frac{\mu^{4}}{4g} .
\end{equation}
\begin{equation}\begin{array}{l}
a=\vert\phi\vert^{\frac{8\pi GV_{0}}{\mu^{2}}}\vert
g\phi^{2}-\mu^{2}\vert^{\frac{\pi G\mu^{2}}{g}-\frac{4\pi
GV_{0}}{\mu^{2}}}\exp(-\pi G\phi^{2}) .
\end{array}\end{equation}
If $\phi\rightarrow 0$ then $a\rightarrow 0$ (for $V_{0}>0$,
 if $V_{0}=0$ then $a\rightarrow const\neq 0$). If $\phi\rightarrow\mu g^{-\frac{1}{2}} $ then $a$ goes
either to 0 or to infinity depending on the value of $V_{0}$
(for $V_{0}=\frac{\mu^{4}}{4g}$ we have $a\rightarrow const\neq
0$) . When $\phi\rightarrow\infty$ then $a\rightarrow 0$.
According to the formula (31) ($B=W=0$) the classical slow roll
time evolution of $\phi$ is determined by
\begin{equation}
-\sqrt{24\pi G}\frac{d\phi}{dt}=\phi(g\phi^{2}-\mu^{2})(V_{0}
+\frac{g}{4}\phi^{4}-\frac{\mu^{2}}{2}\phi^{2})^{-\frac{1}{2}}
\end{equation}
Hence, if $\mu g^{-\frac{1}{2}}\geq \phi\geq 0$ then $\phi$ is
increasing. If $\phi\geq \mu g^{-\frac{1}{2}}$ then $\phi$ is
decreasing to $\mu g^{-\frac{1}{2}}$. It follows that
$\phi(t)\rightarrow \mu g^{-\frac{1}{2}}$ when $t\rightarrow
\infty$. If $V_{0}=\frac{\mu^{4}}{4g}$ then $V+V_{0}\rightarrow 0$
for $t\rightarrow \infty$. Then, the slow-roll condition is
violated for a large time (this remains true for $\mu=0$).
Nevertheless, the stochastic dynamical system (31) still makes
sense. As will be shown in the next section the stochastic
equation with the Starobinsky-Vilenkin noise leads to a
non-integrable stationary probability distribution (if treated as
a system on the whole real line) whereas the system with the
environmental noise has a normalizable stationary distribution.

\section{Stationary probability distribution of the inflaton} The probability distribution $P_{t}(\phi,\phi_{0})$ determines the
probability of an appearance of the universe with given $\phi$ ( or
$a(\phi)$) when initially it had the value $\phi_{0}$.
 The stationary probability $P(\phi)$ is the limit of $P_{t}$ for $t\rightarrow\infty$ \cite{risken}
 (it does not depend on  $\phi_{0}$). In the cosmic time and with
 the Stratonovitch interpretation it can be obtained from the
 requirement $\partial_{t}P=0$ which gives (after an integration
 over $\phi$)\begin{equation}\begin{array}{l}
\frac{\gamma^{2}}{18}\frac{1}{Ha^{\frac{3}{2}}}\partial_{\phi}\frac{1}{Ha^{\frac{3}{2}}}P
+\frac{1}{8\pi^{2}}H^{\frac{3}{2}}\partial_{\phi}H^{\frac{3}{2}}P
+(3H)^{-1}V^{\prime}P =0.\end{array}\end{equation}
For the e-fold time
\begin{equation}\begin{array}{l}
\frac{\gamma^{2}}{18}\frac{1}{H^{\frac{3}{2}}a^{\frac{3}{2}}}\partial_{\phi}\frac{1}{H^{\frac{3}{2}}a^{\frac{3}{2}}}P
+\frac{1}{8\pi^{2}}H\partial_{\phi}H P
+(3H)^{-2}V^{\prime}P =0.\end{array}\end{equation}

 In the
multidimensional case (41) the requirement $\partial_{t}P=0$ is
satisfied if
 \begin{equation}\begin{array}{l}
\frac{\gamma^{2}}{18}\frac{1}{Ha^{\frac{3}{2}}}\partial_{j}\frac{1}{Ha^{\frac{3}{2}}}P
+\frac{1}{8\pi^{2}}H^{\frac{3}{2}}\partial_{j}H^{\frac{3}{2}}P
+(3H)^{-1}\partial_{j}V P=0 .\end{array}\end{equation} This is a
general form of the equation for rotationally invariant stationary
 probability.  If the functions $P$ and $V$ depend
only on $\vert \phi\vert$ then eq.(53) is equivalent to
\begin{equation}\begin{array}{l}
\frac{\gamma^{2}}{18}\frac{1}{Ha^{\frac{3}{2}}}\partial_{\vert\phi\vert}\frac{1}{Ha^{\frac{3}{2}}}P
+\frac{1}{8\pi^{2}}H^{\frac{3}{2}}\partial_{\vert\phi\vert}H^{\frac{3}{2}}P
+(3H)^{-1}P \partial_{\vert\phi\vert}V=0.\end{array}\end{equation}
In our model of diffusion let us consider the simplest cases
first. The stationary solution of eq.(51) without the
Starobinsky-Vilenkin noise is
\begin{equation}\begin{array}{l}
 P=\sqrt{V+V_{0}}\exp(-12\pi
G\int^{\phi}d\phi^{\prime}
(V^{\prime})^{-1}(V+V_{0})\Big)\cr\exp\Big(-\frac{6}{\gamma^{2}}\sqrt{\frac{8\pi
G}{3}}\int d\phi V^{\prime}\sqrt{V+V_{0}}\exp(-24\pi
G\int^{\phi}d\phi^{\prime} (V^{\prime})^{-1}(V+V_{0})\Big),
\end{array}\end{equation} where the exponential factors in eq.(55) come
from the formula for $a^{3}$ (eq.(37)).
 If we assume that $V$ does not grow faster than exponentially and
 is an even function of $\phi$ then for a large $\vert\phi\vert$
\begin{equation}\begin{array}{l}
 P\simeq \sqrt{V+V_{0}}\exp\Big(-12\pi
G\int^{\phi}d\phi^{\prime}
(V^{\prime})^{-1}(V+V_{0})\Big)\end{array}
\end{equation} because the last factor in eq.(55) tends to 1.
If $\gamma=0$ (the environmental noise is absent) then we obtain
the Starobinsky solution (discussed also by Vilenkin
\cite{vilenkin} and Linde \cite{linde}) which in the Stratonovitch
interpretation takes the form

\begin{equation}
P=(V+V_{0})^{-\frac{3}{4}}\exp(\frac{3}{8G^{2}}\frac{1}{V+V_{0}}).
\end{equation}
Ito interpretation gives the factor $(V+V_{0})^{-\frac{3}{2}}$
whereas  e-fold time the factor $(V+V_{0})^{-1}$ in front of the
exponential. The probability distribution (57) is also the
stationary solution in the multifield rotationally symmetric
D-dimensional case (54). Then, for $V\simeq \vert\phi\vert^{n}$ it
fails to be integrable at large $\vert\phi\vert$ if
$\frac{3}{4}n\leq D$ in Stratonovitch case and $n\leq D$ in the
e-fold time (as discussed in \cite{venn}\cite{venn3}). The authors
\cite{venn}\cite{venn3} impose  boundary conditions at large
$\phi$ and discuss whether the dependence on boundary conditions
has consequences on some measurable expectation values. In the
system with an environmental noise and the $\phi^{n}$ potentials
we stabilize the system by the environmental noise instead of
restricting it  by boundary conditions ( when $P$ in eq.(57) is
not integrable). The environmental noise is present in all
physical systems. Its crucial role in equilibration of dynamical
systems is well-known \cite{ruell}.  Its action can really be seen
as a stabilization because some correlation functions will have a
singular or chaotic behaviour at $t\rightarrow \infty$ when
$\gamma=0$ and a smooth behaviour when $\gamma>0$ (the same remark
applies to boundary conditions;some correlation functions with
boundary conditions are discussed  in \cite{venn3} ). If $D=1$ and
$(V+V_{0})=\phi^{n}$ with $n\geq 2$ then $P$ is integrable at
large $\phi$ but not integrable at $\phi=0$. $\phi\simeq 0$ does
not satisfy the slow-roll conditions. We could remove it from the
configuration space imposing boundary conditions at $\phi\simeq
8\pi G$. However, such boundary conditions change the stochastic
equations \cite{ikeda} and the equations for probability
distributions. In general, it is not simple to estimate the
stationary probability distribution. The formula (57) fails to
express a probability distribution ($P$ is not integrable) if $V$
does not fall quickly enough for large $\phi$. This  is the case
with the exponential potential (discussed at eq.(44)) when
$\phi\rightarrow -\infty$ and with the flat potential (45).
Moreover,  $P$ in eq.(57) may be non-integrable at finite $\phi$
if $V+V_{0} =0$ at a certain $\phi_{c}$ as for the natural
inflation (47) and the double well potential (48). The
non-integrability occurs
 in the range of $\phi$ which is outside of the slow-roll regime
determined by the small values of the parameters $\epsilon $ and
$\tilde{\eta}$. We could expect that $P$ of eq.(57) is still an
approximate stationary probability (after an imposition of
boundary conditions) in the slow-roll region and integrable there.
    We shall show that for some potentials
the environmental noise leads to explicit formulas for the
stationary probability without the problems with integrability.
For this reason the introduction of the environmental noise could
be considered as an alternative method of stabilization of the
probability distribution. It allows an extension of the stochastic
equation beyond the slow-roll (or inflation) region.

With $\gamma\neq 0$ for the Stratonovitch equation
(51) we write
\begin{displaymath}
\tilde{P}=H^{-1}a^{-\frac{3}{2}}P .
\end{displaymath}
Then, the equation for $\tilde{P}$ is
\begin{equation}
\frac{\gamma^{2}}{18}H^{-1}a^{-\frac{3}{2}}\partial_{\phi}\tilde{P}+\frac{1}{8\pi^{2}}H^{\frac{3}{2}}\partial_{\phi}(H^{\frac{5}{2}}a^{\frac{3}{2}}\tilde{P})=
-\frac{1}{3}V^{\prime}a^{\frac{3}{2}}\tilde{P}.\end{equation}
Using the formulas for $H$ and for  $a$ (eq.(37)) we obtain
\begin{equation}\begin{array}{l}
\ln\tilde{P}=-6\int d\phi H a^{3}
(\gamma^{2}+\frac{9}{4\pi^{2}}H^{5}a^{3})^{-1}\cr\Big(V^{\prime}+\frac{10}{3}G^{2}(V+V_{0})V^{\prime}
-32\pi G^{3}(V+V_{0})^{3} (V^{\prime})^{-1}\Big).\end{array}
\end{equation}
For most potentials $a\rightarrow 0$ for $\phi\rightarrow \infty$
then $a^{3}H^{5}\rightarrow 0$ and we get from eq.(59) $P\simeq
Ha^{\frac{3}{2}}$ coinciding with the formula (56). It leads to
stationary solutions which are integrable  at large $\phi$. As an
explicit example we consider the chaotic inflation potential
$\phi^{n}$ with $V_{0}=0$. Then,  the formula for $a$ (43) gives
for a large $\phi$ (small $a$)
\begin{equation}
P=\vert\phi\vert^{\frac{n}{2}}\exp(-6\pi Gn^{-1}\phi^{2}) .
\end{equation} At small $\phi$ and $V_{0}=0$ we have $a^{3} H^{5}\rightarrow 0$ . So, the formulae (56)
and (60) are   applicable  for small as well as large $\phi$.The
formula (60) (which holds true for multidimensional fields $\phi$)
describes in statistics the gamma distribution ( more precisely
the $\chi^{2} $ distribution).
 The Starobinsky formula (57)
(obtained without the thermal noise) gives $P$ which is not
integrable at $\phi=0$. If $V_{0}\neq 0$ for $\phi^{n}$
potential then  $a(\phi)\rightarrow \infty$ for $\phi\rightarrow
0$ and $a^{3}H^{5}\rightarrow \infty$. Then, the formula (57)
applies,  but with $V_{0}>0$, hence it is integrable.

In general, when $a^{3}H^{5}\rightarrow \infty$ then
\begin{equation}
\ln\tilde{P}\simeq -\frac{8\pi^{2}}{3}\int d\phi H^{-4}
\Big(V^{\prime}+\frac{10}{3}G^{2}(V+V_{0})V^{\prime} -32\pi
G^{3}(V+V_{0})^{3} (V^{\prime})^{-1}\Big)
\end{equation}
We can calculate the rhs of eq.(61) and convince ourselves that in
this case we reach the Starobinsky formula (57).

We could set
\begin{displaymath}
\hat{P}=H^{\frac{3}{2}}P
\end{displaymath}
then the equation for $\hat{P}$ reads
\begin{displaymath}
\frac{\gamma^{2}}{18}H^{-1}a^{-\frac{3}{2}}\partial_{\phi}(H^{-\frac{5}{2}}a^{-\frac{3}{2}}\hat{P})
+\frac{1}{8\pi^{2}}H^{\frac{3}{2}}\partial_{\phi}\hat{P}=
-\frac{1}{3}H^{-\frac{5}{2}}V^{\prime}\hat{P}
\end{displaymath}
When we calculate the derivatives of $H$ and $a$ then we obtain
\begin{equation}\begin{array}{l}
\Big(\gamma^{2}+\frac{9}{4\pi^{2}}a^{3}H^{5}\Big)\partial_{\phi}\ln\hat{P}\cr=
-6Ha^{3}\Big(V^{\prime}+2\pi
G\gamma^{2}a^{-3}H^{-1}(V+V_{0})(V^{\prime})^{-1}-\gamma^{2}\frac{5}{24}H^{-1}(V+V_{0})^{-1}V^{\prime}\Big)
\end{array}\end{equation} From eq.(62) we can also see
that if $a\rightarrow 0$ for a large $\phi$ then
\begin{equation}
P\simeq Ha^{\frac{3}{2}}
\end{equation}
 in agreement with eq.(56).

In the e-fold equation (52) we write
\begin{equation}
P=H^{\frac{3}{2}}a^{\frac{3}{2}}P_{e}
\end{equation}
Then, the equation for $P_{e}$ reads
\begin{equation}\begin{array}{l}
-\partial_{\phi}\ln
P_{e}=-6\Big(\gamma^{2}+\frac{9}{4\pi^{2}}a^{3}H^{5}\Big)^{-1} \cr
Ha^{3}\Big(V^{\prime}+\frac{10}{3}G^{2}(V+V_{0})V^{\prime} -32\pi
G^{3}(V+V_{0})^{3} (V^{\prime})^{-1}\Big)
\end{array}
\end{equation}
 There remains to explore the
formula (62) for large $a$ in various models. If
$a^{3}V^{\frac{5}{2}}\rightarrow \infty$ then the terms
independent of $a$ in eq.(62) can be omitted and we get the
Starobinsky formula (57) as we did on the basis of eq.(61). As
expected eqs.(59) and (62) lead to the same formula for $P$ but
they may be applied for direct estimates in different asymptotic
regions.

 Let us still investigate the stationary probability distributions for the remaining potentials of sec.5.
 For the flat potential (45) the Starobinsky formula (57) is not
integrable at  large $\phi$. However, if $K>L\geq 0$ then
$a(\phi)\rightarrow 0$ and $a^{3 }H^{5}\rightarrow 0$ . Hence, the
distribution (56) which is integrable applies at large $\phi$.
However,
 if $L>K>0 $ then  $a^{3}H^{5}\rightarrow \infty$. Then, for a large $\phi$ the Starobinsky formula  (57)
   arises as a limit which  is not integrable.
This example shows that the thermal noise stabilization is not
universal (it seems to apply if $a$ does not grow with $\phi$). In
the exponential model $a\rightarrow 0$ if $\phi\rightarrow
+\infty$. Then,  $a^{3}H^{5}\rightarrow 0$ and eq.(56) applies.
Hence, $P$ is integrable at large positive $\phi$. $a(\phi)$ tends
to infinity if $\phi\rightarrow -\infty$. If $V_{0}>0$ then
$a^{3}H^{5}\rightarrow \infty$. The asymptotic behaviour of $P$ is
determined by the distribution (57) which is not integrable at
large negative $\phi$.
 In the double
well model if $V_{0}=\frac{\mu^{4}}{4g}$ then the Starobinsky
distribution (57) would not be integrable
 because of the singularity at $V+V_{0}\simeq 0$. However,
$a^{3}H^{5}$  does not tend to infinity. Hence, eq.(62) applies
which leads to an integrable stationary distribution (approximated
by eq.(56)). The Starobinsky solution (57) for the natural
inflation (47) behaves for a small $\phi$ as \begin{displaymath}
P\simeq\vert\phi\vert^{-\frac{3}{2}}\exp(\frac{16}{3G^{2}\phi^{2}})
\end{displaymath}
Hence, it is not integrable. The environmental noise changes the
small $\phi$ behaviour (as follows from eqs.(59) and(62)) (because
$a^{3}V^{\frac{5}{2}}$ is negligible in comparison to $\gamma^{2}$
for small $\phi$). From eq.(62)  for a small $\phi$

\begin{displaymath}
P\simeq \vert\phi\vert
\end{displaymath}
Hence, $P$ as determined by eq. (62) is a well-defined probability
measure on the interval $-\pi\geq \phi\leq \pi$  .

Summarizing, if $a^{3}H^{5}$ tends to zero we obtain the formula
(56) (either for small or large $\phi$) with an integrable $P$. If
$a^{3}H^{5}$ tends to infinity then asymptotically the behaviour
of $P$ is determined by the Starobinsky formula (57) which may
fail the integrability condition. In the last case the boundary
conditions would be necessary in order to obtain an integrable
stationary distribution.
\section{Summary and conclusions} The Starobinsky-Vilenkin stochastic equation
is a useful tool in a study of an expanding universe at an early
quantum stage. Recently, it has been applied to the calculation of
correlation functions of e-folds and the power spectrum
\cite{starven}\cite{venn}\cite{venn3}\cite{wands}\cite{fujita}.
For some potentials the stationary probability distribution is not
integrable when applied to the whole range of field values. For a
single inflaton field the non-integrability occurs for polynomial large field values only if $ n<2$  (e-fold time). For small field values the non-integrability is
  beyond the
range of the slow-roll regime when the inflation stops. In the
case of multidimensional inflaton field the non-integrability
appears  at large field values  when the standard slow roll
conditions are satisfied \cite{venn}\cite{venn3}. A way out of
this situation (as suggested in \cite{venn}\cite{venn3}) is to
apply the stochastic equations in a region with boundaries
excluding the forbidden regions. However, such a procedure
modifies the stochastic equations and does not lead to explicit
estimates on the stationary probability distribution. Our work
suggests another approach which to some extent is implicit also in
the warm inflation. When the cold inflation stops the
environmental (thermal) noise becomes relevant. In \cite{venn} the
appearance of non-integrability (infinite inflation) is  compared
to a phase transition in solid state physics. It is known
\cite{ruell} that close to the critical point the environment  in
the form of noise or boundaries (as, e.g., in the Ising model)
becomes relevant. This is exactly what happens in the models
discussed in this paper. When the field value goes beyond the
inflation range the boundary conditions become relevant (in the
framework of \cite{venn}\cite{venn3}) or in our alternative
formulation the environmental noise begins to play its role. The
stationary probability depends substantially on the noise although
correlation functions of some observables may only marginally
depend of it. We think that the environmental noise present in any
(not idealized) physical system is a better tool than some
arbitrary boundary conditions. The thermal noise accompanies
classical as well as quantum field theories when an interaction
with an environment is taken into account. The environment may
consist of any degrees of freedom which  undergo an averaging
procedure. We have shown that the addition of the thermal noise
has a stabilizing effect on quantum fluctuations. In sec.4 we have
shown that if $a^{3}H^{5}$ stays bounded then the thermal noise
repairs $P$ even beyond the slow-roll regime.The non-integrability
of the stationary solution excludes it from calculation of
expectation values. In such a case we cannot expect a finite
result for the power spectrum at $t\rightarrow \infty$ (it is
calculated in \cite{starven}\cite{fujita} at finite time). The
standard estimates of the slow roll parameters $\epsilon \simeq
(V^{-1}V^{\prime})^{2}$ and $\tilde{\eta}\simeq
V^{-1}V^{\prime\prime}$ restrict the values of fields in the form
of boundaries but they could be replaced by expectation values
which could be small because the forbidden values of the fields
would have small probabilities with the presence of the thermal
noise.

\end{document}